\documentclass[journal=jcisd8,manuscript=article]{achemso}

\usepackage[T1]{fontenc}
\usepackage{amsmath}
\usepackage{amssymb}
\usepackage{booktabs}
\usepackage{graphicx}
\usepackage{xcolor}
\usepackage{xr}
\usepackage{pdfpages}
\newcommand{\AP}{\mathrm{AP}}
\newcommand{\EF}{\mathrm{EF}}

\author{Kairi Furui}
\affiliation[ScienceTokyo]{Department of Computer Science, School of Computing, Institute of Science Tokyo, Yokohama 226-8501, Japan}

\author{Masahito Ohue}
\email{ohue@comp.isct.ac.jp}
\phone{+81 (0)45 924 5522}
\fax{+81 (0)45 924 5523}
\affiliation[ScienceTokyo]{Department of Computer Science, School of Computing, Institute of Science Tokyo, Yokohama 226-8501, Japan}

\title[Adapting Boltz-2 for hit discovery]
  {Adapting Boltz-2 with limited experimental activity data improves early enrichment in virtual screening}

\abbreviations{HTS, AP, EF, BEDROC}

\newcommand{\TargetHeader}{Assay ID}
\newcommand{\RateHeader}{Active rate}
\newcommand{\TrainingActiveHeader}{Training actives}
\newcommand{\MetricHeader}{Metric}
\newcommand{\RatioHeader}{Geo. mean ratio}
\newcommand{\ImprovedHeader}{Improved}
\newcommand{\SignHeader}{Sign $p$}

\ifdefined\DeclareUnicodeCharacter
\DeclareUnicodeCharacter{0394}{\ensuremath{\Delta}}
\fi
\begin{document}

\begin{abstract}
Virtual screening aims to prioritize active compounds from large chemical libraries within a limited experimental budget.
When applying Boltz-2 to virtual screening, a key challenge is how to use limited experimental data from the target assay to improve the prioritization of active compounds.
We investigated whether fine-tuning the Boltz-2 affinity heads with a small number of binary activity labels could improve early enrichment of active compounds in hit discovery.
We compared fine-tuning with 40--300 labels in a retrospective evaluation on eight MF-PCBA targets.
With 300 activity measurements, fine-tuning increased the number of actives in the top 1\% by a geometric mean of 1.77-fold across the eight targets and improved average precision (AP) by 2.14-fold relative to the control without fine-tuning.
We also investigated whether rescoring a subset of candidates could retain the improvement in hit recovery by reranking only the top-ranked Boltz-2 candidates with the fine-tuned head.
Restricting rescoring to approximately 10\% of the evaluation set retained hit recovery comparable to full rescoring.
These findings show that affinity-head fine-tuning with limited activity labels improves early enrichment with Boltz-2 and that this benefit can be retained when rescoring a restricted set of candidates.
\end{abstract}

\section{Introduction}

Early-stage hit discovery requires selecting compounds for follow-up from large libraries.
High-throughput screening (HTS) can measure many compounds, but procurement, assays, and activity confirmation require time and resources.
In virtual screening, early enrichment near the top of the ranking is essential for recovering active compounds within a limited measurement budget~\cite{Truchon2007-ff}.

When experimental data are available for the target assay, training an assay-specific activity predictor from molecular descriptors is a standard approach to candidate selection~\cite{Tran-Nguyen2023-ak,Feng2024-vu,Feng2025-fd}.
Extended-connectivity fingerprints (ECFPs) and pretrained molecular representations provide inputs for such predictors~\cite{Rogers2010-kp}.
When only approximately 100--300 measurements are available at the beginning of hit discovery, a central question is how to incorporate these active/inactive observations into prediction.

Structure-based virtual screening uses information about target--ligand interactions to select candidates.
AutoDock Vina and GNINA explore binding modes and score candidates using empirical functions and deep learning, respectively~\cite{Trott2010-aq,Eberhardt2021-jr,McNutt2021-nq}.
Co-folding models predict complex structures from proteins and ligands, and Boltz-2 also integrates binding affinity prediction~\cite{Abramson2024-mf,Wohlwend2025-rn,Passaro2025-tw}.
For Boltz-2-based hit discovery, Cecchini and Sinenka evaluated rescoring and inference settings across multiple targets, while BoltzMol-1 examined prospective compound selection~\cite{Cecchini2026-pq,Getz2026-uc}.
Adapting these pretrained models to experimental observations from the target assay offers another way to use limited data, alongside training a new predictor from molecular descriptors.

To enable such adaptation, Amini et al. released an implementation for fine-tuning the affinity prediction components of Boltz-2 using project-specific experimental data~\cite{Amini2026-ee}.
Their work examined potency prediction for lead optimization through a retrospective multi-target benchmark and a single-target study using internal data.
Here, we apply this fine-tuning approach to hit discovery using binary HTS activity measurements and evaluate early enrichment when selecting follow-up candidates from libraries dominated by inactive compounds.

Training compounds are selected from the top of the initial screening ranking, where candidates for experimental follow-up are drawn.
This selection provides both active compounds and compounds found to be inactive despite receiving high predicted scores.
We therefore evaluate fine-tuning using activity measurements for compounds selected from the top of the initial ranking.

We evaluated fine-tuning the Boltz-2 affinity head with 40, 100, and 300 binary activity measurements on eight MF-PCBA targets~\cite{Buterez2023-ux}.
First, we quantified gains in early enrichment by comparing performance before and after fine-tuning.
Next, we compared head updates with DrugCLIP projection-head fine-tuning and LightGBM trained on identical measurements to assess performance relative to learning from molecular descriptors or cached representations~\cite{Ke2017-dr}.
Finally, we restricted rescoring to the top-ranked Boltz-2 candidates to determine how broadly fine-tuning must be applied to retain its improvement in hit recovery.

\section{Materials and Methods}

We evaluated fine-tuning in terms of label budget, learning approaches using identical labels, and the scope of rescoring.
The following sections describe label acquisition and evaluation-set construction, model training and scoring conditions, and the assessment of early enrichment and hit recovery.

\subsection{Data Set and Label Acquisition}

We used the eight-target MF-PCBA compound sets assembled in our previous study, Boltzina~\cite{Buterez2023-ux,Furui2025-ux}.
These correspond to eight of the 10 assays used in the Boltz-2 evaluation by Passaro et al.
The remaining two assays were excluded in the previous study because the binding site could not be specified or many active compounds contained more than 60 heavy atoms~\cite{Passaro2025-tw,Furui2025-ux}.
We retained the distributed compounds and binary labels and included compounds for which prediction scores and representations were available.
Table~\ref{tab:targets} summarizes the evaluation sets and the numbers of actives in the training sets.

Training compounds were selected from the top of the binding-probability ranking produced by standard Boltz-2 inference, with label budgets $N$ of 40, 100, and 300.
The training sets were nested, expanding down the initial ranking.
We excluded the top 300 compounds from evaluation in every condition, yielding common evaluation sets of 49,582--49,695 compounds per target.
The evaluation population and active fraction were identical across label budgets.
Using existing measurement labels, we retrospectively evaluated a setting in which activity measurements for compounds selected in an initial screen are used for training.

\begin{table}[!htbp]
\centering
\caption{MF-PCBA evaluation and training sets. Assays are identified by their PubChem identifiers. $n_{\mathrm{eval}}$ and $n_{\mathrm{active}}$ denote the number of compounds and actives in the evaluation set after excluding the top 300 compounds, which contain all training candidates. The final three columns report the numbers of training actives at each label budget.}
\label{tab:targets}
\small
\begin{tabular}{lrrrrrr}
\toprule
\TargetHeader & $n_{\mathrm{eval}}$ & $n_{\mathrm{active}}$ & \RateHeader & \multicolumn{3}{c}{\TrainingActiveHeader} \\
\cmidrule(lr){5-7}
 & & & (\%) & $N=40$ & $N=100$ & $N=300$ \\
\midrule
588689 & 49,685 & 396 & 0.80 & 16 & 32 & 90 \\
540297-493091 & 49,685 & 739 & 1.49 & 8 & 16 & 43 \\
434954-2097 & 49,615 & 415 & 0.84 & 24 & 52 & 107 \\
463203-2650 & 49,582 & 542 & 1.09 & 13 & 29 & 70 \\
493248-485317 & 49,680 & 953 & 1.92 & 1 & 9 & 23 \\
504329 & 49,682 & 438 & 0.88 & 4 & 11 & 28 \\
624273-588549 & 49,687 & 150 & 0.30 & 4 & 4 & 9 \\
1053173-743445 & 49,695 & 134 & 0.27 & 0 & 5 & 10 \\
\bottomrule
\end{tabular}

\end{table}

\subsection{Boltz-2 Inputs and Scoring Conditions}

Boltz-2 comprises input embeddings, a trunk, a structure module, and modules for affinity prediction~\cite{Passaro2025-tw}.
We refer to each of the two affinity prediction modules as an affinity head.
The final screening score was the arithmetic mean of the predicted binding probabilities output by the two heads.
Protein inputs were the target-specific construct sequences prepared in the previous study; neither known complex structures nor docking-generated poses were provided.
Multiple sequence alignments (MSAs) were generated for each target using the Boltz MSA server functionality and shared across its compounds.

\begin{figure}[!htbp]
\centering
\includegraphics[width=\linewidth]{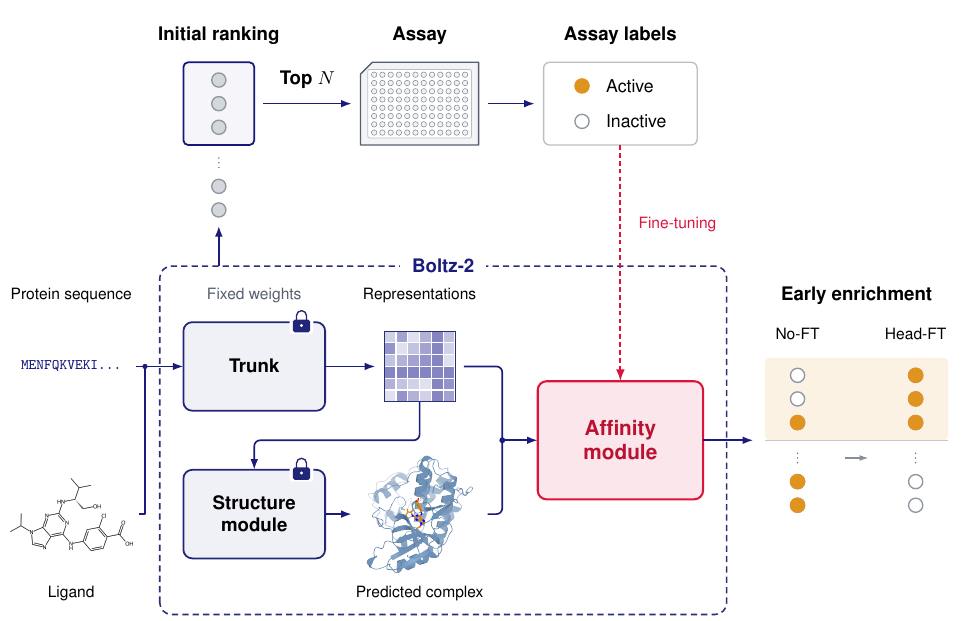}
\caption{Overview of Boltz-2 fine-tuning and screening using limited assay measurements.}
\label{fig:overview}
\end{figure}

We denote the control without fine-tuning as No-FT and the fine-tuned condition as head-FT.
At evaluation, No-FT and head-FT shared identical poses, identical representations generated with the affinity checkpoint, and the same scoring pathway.
Outputs of the trunk and structure module were stored and reused for scoring.
Standard Boltz-2 inference reselects a pose during the affinity prediction stage, whereas our evaluation used poses obtained before that stage.
We therefore distinguished the initial ranking used to select training compounds from the No-FT ranking used to measure the fine-tuning effect.
Figure~\ref{fig:overview} shows the Boltz-2 architecture and the components that are fine-tuned.

\subsection{Affinity-Head Fine-Tuning}

We used a derivative of Amini et al.'s implementation adapted to binary labels and stored representations~\cite{Amini2026-ee}.
Starting from the pretrained affinity checkpoint, we updated both affinity heads using binary focal loss ($\alpha=0.25$, $\gamma=2$).
The trunk and structure module weights were fixed, and inputs to the head were detached from the computational graph.

Training also loaded stored poses and pair representations and constructed input features corresponding to ligand-centered crops.
The maximum numbers of tokens, atoms, and MSA sequences were 512, 4608, and 2048, respectively.
Input featurization was configured to apply pocket conditioning with probability 0.3 and cutoffs of 4--20~\AA{}.

Training was fixed at five epochs, with a batch size of one, gradient accumulation over four steps, a maximum learning rate of $2\times10^{-5}$, and two warmup steps.
Each condition was trained once, and the final-epoch checkpoint was evaluated.
The last 20 compounds in the initial ranking of the evaluation set were used for validation during training.
Training and scoring used a single NVIDIA H100 GPU.

\subsection{Supervised Baselines Using Identical Labels}

LightGBM and DrugCLIP used the same training compounds and binary labels as head-FT and were scored on the common evaluation set.
For both methods, hyperparameters were selected to maximize mean AP in stratified three-fold cross-validation within the training set.
Boltz-2 head-FT used the fixed settings described above.

Conditions with fewer than three active or three inactive training compounds could not support stratified three-fold cross-validation and were excluded from the comparison for all methods.
All methods used the same targets in this comparison: six at $N=40$ and eight at $N=100$ and 300.
Search ranges and excluded targets are listed in the Supporting Information.

\noindent\textbf{LightGBM.}
We used five feature settings: trunk representations, CheMeleon, ECFP, and concatenations of either CheMeleon or ECFP with trunk representations~\cite{Ke2017-dr}.
Trunk representations were 128-dimensional features obtained by averaging Boltz-2 pair representations over ligand--receptor token pairs in both directions and off-diagonal ligand--ligand token pairs.
CheMeleon provided 2048-dimensional molecular embeddings from a pretrained model~\cite{Burns2025-mq}, and ECFP used 2048-bit Morgan fingerprints with a radius of two~\cite{Rogers2010-kp}.
Both concatenated feature sets had 2176 dimensions.

We searched 27 combinations of the number of trees, number of leaves, and learning rate.
Predicted probabilities were averaged over five models trained independently with the selected configuration.

\noindent\textbf{DrugCLIP.}
We evaluated zero-shot scoring with the publicly available pretrained weights of DrugCLIP, which embeds pockets and molecules in a shared space~\cite{Gao2023-iw}.
To adapt the representations learned through contrastive training in the original study to target-specific activity classification, we introduced projection-head fine-tuning with binary focal loss.
Using the predicted complexes of target proteins and reference ligands from our previous study, we selected residues containing a heavy atom within 6~\AA{} of a ligand heavy atom and included all heavy atoms of these residues in the pocket~\cite{Furui2025-ux}.
Each compound was represented by one conformer generated with RDKit ETKDGv3 and MMFF.

Fine-tuning froze the molecular and pocket encoders and updated both projection heads and the score scale parameter.
We searched 18 combinations of the learning rate, number of updates, and L2-SP regularization coefficient.
Fine-tuning was repeated five times per condition, and metrics were averaged within each target before aggregation across targets.

\subsection{Two-Stage Candidate Selection}

If rescoring a subset of candidates retains the improvement in hit recovery, fewer additional evaluations are needed after the initial screen.
We therefore ranked the common evaluation set using Boltz-2 before fine-tuning, denoted No-FT, rescored its top $M$ compounds with head-FT, and selected the top $k$ compounds in the resulting ranking.
We evaluated $M$ of 500, 1,000, 2,500, 5,000, 10,000, and 25,000, together with head-FT rescoring of the entire evaluation set.
The final candidate counts $k$ were 100 and 500.
Recovery was defined as the number of actives among the $k$ compounds selected from the common evaluation set after excluding the training candidates.

The computational comparison concerned the number of head-FT evaluations after an existing No-FT ranking was available.
Initial Boltz-2 inference, representation storage, and fine-tuning were preprocessing steps shared across all rescoring conditions.

\subsection{Evaluation Metrics and Statistics}

\noindent\textbf{Average precision (AP).}
AP was calculated by weighting precision by the increments in recall as the score threshold decreases:
\begin{equation}
\AP=\sum_j (R_j-R_{j-1})P_j,
\label{eq:ap}
\end{equation}
where $P_j$ and $R_j$ are the precision and recall at threshold $j$.

\noindent\textbf{Enrichment in the top 1\% (EF@1\%).}
For an evaluation set containing $n$ compounds and $a$ actives, with $h_m$ actives among the top $m$ compounds, enrichment was defined as
\begin{equation}
\EF_{1\%}=\frac{h_m/m}{a/n},\qquad m=\max\{1,\operatorname{round}(0.01n)\}.
\label{eq:ef}
\end{equation}

\noindent\textbf{BEDROC.}
We used BEDROC with $\alpha=20$ to emphasize the highest-ranked compounds~\cite{Truchon2007-ff}.

For comparisons between conditions, we calculated the metric ratio for each target and exponentiated the mean log ratio to obtain the geometric-mean ratio.
Absolute performance was summarized by arithmetic means across targets. Recovery comparisons used the mean of within-target differences and its 95\% confidence interval.
The 95\% confidence intervals in the figures and tables were $t$ intervals with targets as the analysis units; intervals for ratios were calculated in log space and then exponentiated.

We used a two-sided sign test for the number of improved targets and a two-sided Wilcoxon signed-rank test on per-target log ratios.
The sign test treated ties as non-improvements and retained them in the denominator. For EF@1\% at $N=100$, one target was tied; therefore, the reported improvement count of $7/8$ consists of seven improved targets and one tie.

\subsection{Analyses of Internal Representations and Chemical Structure}

We examined the association between changes in internal representations and rankings after fine-tuning at $N=300$.
For each target, we sampled 512 compounds uniformly without replacement from the common evaluation set, independently of their activity labels.
Using identical inputs, we compared the 384-dimensional features after the multilayer perceptron (MLP) before and after fine-tuning, separately for each head.
Rank improvement was defined as the No-FT rank minus the head-FT rank within the complete evaluation set.
Feature displacement was defined as the L2 norm of the difference between the feature vectors before and after fine-tuning, divided by the L2 norm of the feature vector before fine-tuning.
Procedures for calculating rank correlations are provided in the Supporting Information.

Chemical scaffolds were defined using Bemis--Murcko scaffolds; evaluation compounds whose scaffolds were absent from all training compounds at a given label budget formed the unseen-scaffold stratum.
Acyclic compounds were treated separately, and scaffolds with generalized atom and bond types were also evaluated.
Chemical similarity was calculated using the Tanimoto coefficient for Morgan fingerprints with radius 2 and 2048 bits.
For the nearest-active method, each evaluation compound was ranked by its maximum Morgan Tanimoto similarity to an active compound in the training set.
Performance metrics were calculated within each stratum, and methods were compared on the same targets containing at least 10 actives in that stratum.
Top-100 recovery was counted from rankings of the complete evaluation sets, with score ties resolved by saved evaluation-ID order.

\section{Results and Discussion}

\subsection{Label Budget and Early Enrichment}

\begin{figure}[!htbp]
\centering
\includegraphics[width=\linewidth]{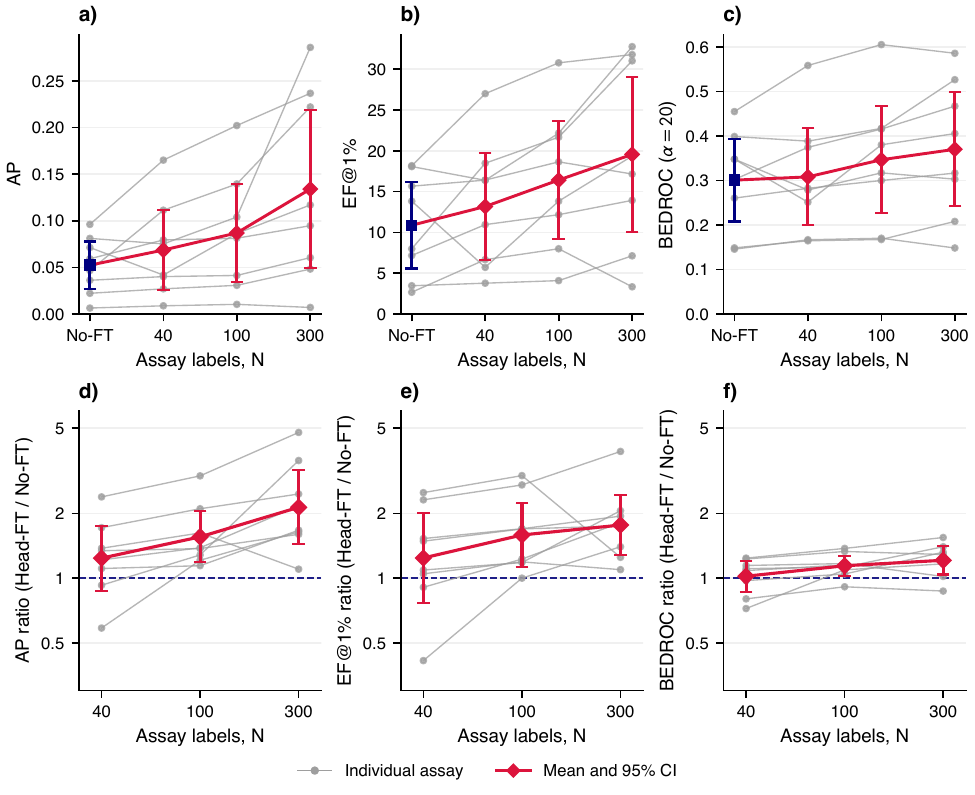}
\caption{Head-FT performance across label budgets. (a--c) Absolute AP, EF@1\%, and BEDROC; (d--f) ratios to No-FT. Gray points and lines show values and connections for eight targets; large symbols show arithmetic means in the top row and geometric means in the bottom row. Error bars indicate 95\% confidence intervals across targets, calculated in log space for ratios. The bottom row uses logarithmic scales, with horizontal dashed lines at a ratio of one.}
\label{fig:learning}
\end{figure}

We examined whether fine-tuning with limited measurements improves early enrichment and how the effect depends on the number of training measurements.
We fine-tuned the affinity heads using labels for the top $N=40$, 100, or 300 compounds in the initial ranking and compared them with No-FT.
Figure~\ref{fig:learning} shows early enrichment across label budgets, and Table~\ref{tab:main} reports absolute performance and improvement ratios relative to No-FT.
The geometric-mean AP ratios to No-FT were 1.239, 1.556, and 2.137 at $N=40$, 100, and 300, respectively.
AP improved for all eight targets at both $N=100$ and $N=300$ (two-sided sign and Wilcoxon tests, both $p=0.0078$).
The geometric-mean EF@1\% ratios also increased from 1.241 to 1.588 and 1.766.
At $N=40$, however, neither test yielded a $p$ value below 0.05 for any of the three metrics.
The ranking gains therefore extended beyond AP to enrichment of actives among the top-ranked candidates for follow-up.

\begin{table}[!htbp]
\centering
\caption{Screening performance and improvement relative to No-FT at each label budget. The $p$ values are from unadjusted two-sided tests.}
\label{tab:main}
\small
\setlength{\tabcolsep}{3pt}
\begin{tabular}{llrrrrrr}
\toprule
\MetricHeader & $N$ & No-FT & Head-FT & \RatioHeader & \ImprovedHeader & \SignHeader & Wilcoxon $p$ \\
\midrule
AP & 40 & 0.052 & 0.069 & 1.239 [0.875, 1.754] & 6/8 & 0.2891 & 0.1484 \\
AP & 100 & 0.052 & 0.087 & 1.556 [1.185, 2.044] & 8/8 & 0.0078 & 0.0078 \\
AP & 300 & 0.052 & 0.134 & 2.137 [1.438, 3.174] & 8/8 & 0.0078 & 0.0078 \\
\addlinespace
EF@1\% & 40 & 10.88 & 13.18 & 1.241 [0.768, 2.006] & 6/8 & 0.2891 & 0.3125 \\
EF@1\% & 100 & 10.88 & 16.43 & 1.588 [1.131, 2.229] & 7/8 & 0.0703 & 0.0156 \\
EF@1\% & 300 & 10.88 & 19.57 & 1.766 [1.276, 2.444] & 8/8 & 0.0078 & 0.0078 \\
\addlinespace
BEDROC & 40 & 0.301 & 0.308 & 1.021 [0.865, 1.205] & 5/8 & 0.7266 & 0.8438 \\
BEDROC & 100 & 0.301 & 0.347 & 1.142 [1.024, 1.274] & 7/8 & 0.0703 & 0.0391 \\
BEDROC & 300 & 0.301 & 0.370 & 1.211 [1.040, 1.410] & 7/8 & 0.0703 & 0.0234 \\
\bottomrule
\end{tabular}

\end{table}

Figure~\ref{fig:learning} shows that the AP ratios at $N=300$ ranged from 1.10 to 4.76.
Because gains varied at the same measurement budget, Table S1 of the Supporting Information provides per-target absolute performance alongside the aggregate results.
Incorporating 100--300 measurements into the existing predictor thus adapted the initial library ranking to the target assay and increased the representation of actives among candidates for follow-up.

\subsection{Comparison with Supervised Models Using Identical Labels}

\begin{figure}[!htbp]
\centering
\includegraphics[width=\linewidth]{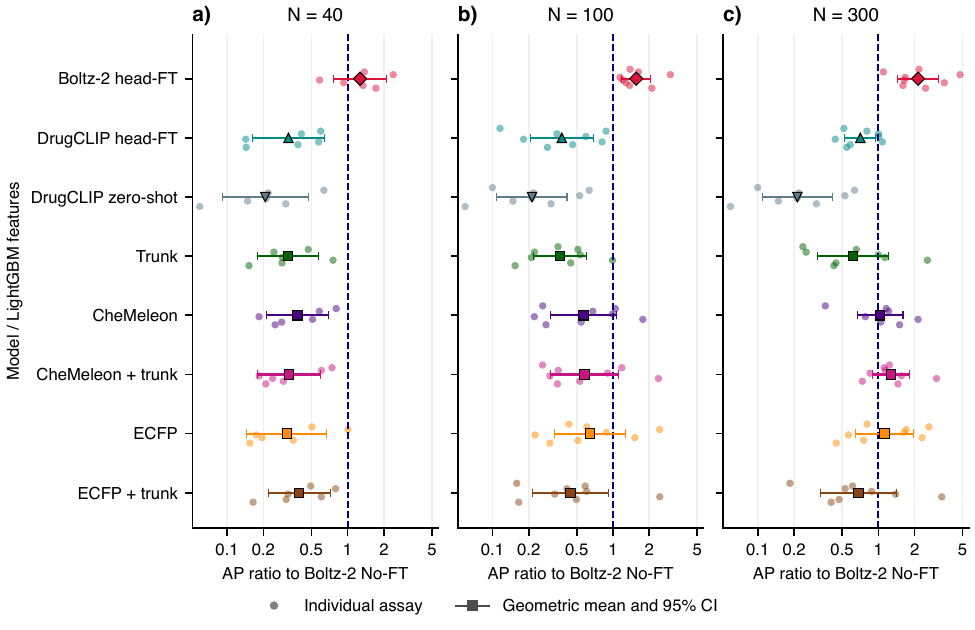}
\caption{AP comparison of Boltz-2 head-FT, DrugCLIP head-FT, and LightGBM using identical labels. The horizontal axis shows AP ratios to Boltz-2 No-FT; Trunk, CheMeleon, and ECFP denote LightGBM features. Small points show per-target values, large symbols show geometric means, and error bars indicate 95\% confidence intervals based on log ratios. All methods aggregate six targets at $N=40$ and eight at $N=100$ and 300.}
\label{fig:baselines}
\end{figure}

We examined how fine-tuning the Boltz-2 affinity heads compares with other supervised models as a way to learn from the available measurements.
We compared AP across models and feature sets using the same training compounds and labels.
Figure~\ref{fig:baselines} compares AP for Boltz-2 head-FT, DrugCLIP projection-head fine-tuning, and LightGBM trained on identical labels.
Boltz-2 head-FT achieved higher aggregate AP than all LightGBM feature settings.
At $N=300$, concatenated CheMeleon and trunk representations gave the highest aggregate AP among these settings, with an AP ratio to No-FT of 1.27, below the 2.14 achieved by head-FT.
DrugCLIP projection-head fine-tuning also improved aggregate AP over zero-shot scoring, but Boltz-2 head-FT achieved higher AP on every target included in the comparison.

\subsection{Association between Feature Displacement and Rank Decline}

We examined whether changes in internal representations during fine-tuning were associated with compounds moving up or down the ranking.
Using models fine-tuned at $N=300$, we compared the post-MLP feature vectors before and after fine-tuning for 512 compounds sampled uniformly from each target's evaluation set, separately for each head.
Feature displacement was measured as the L2 norm of the difference between the two vectors divided by the L2 norm of the vector before fine-tuning.
Ranks were calculated within the complete evaluation set using the mean predicted binding probability from the two heads. Rank improvement was defined as the No-FT rank minus the head-FT rank, with positive values indicating an improvement and negative values indicating a decline.

Feature displacement and rank improvement were negatively correlated for every target in both heads (Table~\ref{tab:head_changes}).
The median Spearman correlations across targets were $-0.575$ for head 1 and $-0.657$ for head 2; partial rank correlations adjusted for No-FT rank and the feature norm before fine-tuning were $-0.577$ and $-0.664$, respectively.
Thus, compounds with larger changes in internal representations tended to move down the ranking after fine-tuning.
Of the 4,096 uniformly sampled compounds, 4,062 were inactive, and the same association was observed when the analysis was restricted to inactive compounds (Section S3 of the Supporting Information).
Because training compounds were selected from the top of the initial ranking, fine-tuning also used labels from high-scoring inactive compounds.
This result is consistent with the interpretation that updates informed by these labels contributed to changes in internal representations and lower ranks for overestimated compounds.

\begin{table}[!htbp]
\centering
\caption{Correlations between post-MLP feature displacement and rank improvement ($N=300$, 512 compounds per target). $\rho$ denotes Spearman correlation; $\rho_{\mathrm{adj}}$ denotes partial rank correlation adjusted for No-FT rank and the feature norm before fine-tuning.}
\label{tab:head_changes}
\small
\begin{tabular}{lrrrr}
\toprule
 & \multicolumn{2}{c}{Head 1} & \multicolumn{2}{c}{Head 2} \\
\cmidrule(lr){2-3}\cmidrule(lr){4-5}
Assay ID & $\rho$ & $\rho_{\mathrm{adj}}$ & $\rho$ & $\rho_{\mathrm{adj}}$ \\
\midrule
1053173-743445 & -0.573 & -0.590 & -0.643 & -0.735 \\
434954-2097 & -0.538 & -0.564 & -0.629 & -0.639 \\
463203-2650 & -0.763 & -0.765 & -0.685 & -0.602 \\
493248-485317 & -0.632 & -0.687 & -0.455 & -0.397 \\
504329 & -0.500 & -0.423 & -0.528 & -0.477 \\
540297-493091 & -0.576 & -0.561 & -0.671 & -0.690 \\
588689 & -0.792 & -0.786 & -0.711 & -0.725 \\
624273-588549 & -0.270 & -0.220 & -0.745 & -0.715 \\
\bottomrule
\end{tabular}

\end{table}

\subsection{Rescoring Budget and Hit Recovery}

\begin{figure}[!htbp]
\centering
\includegraphics[width=\linewidth]{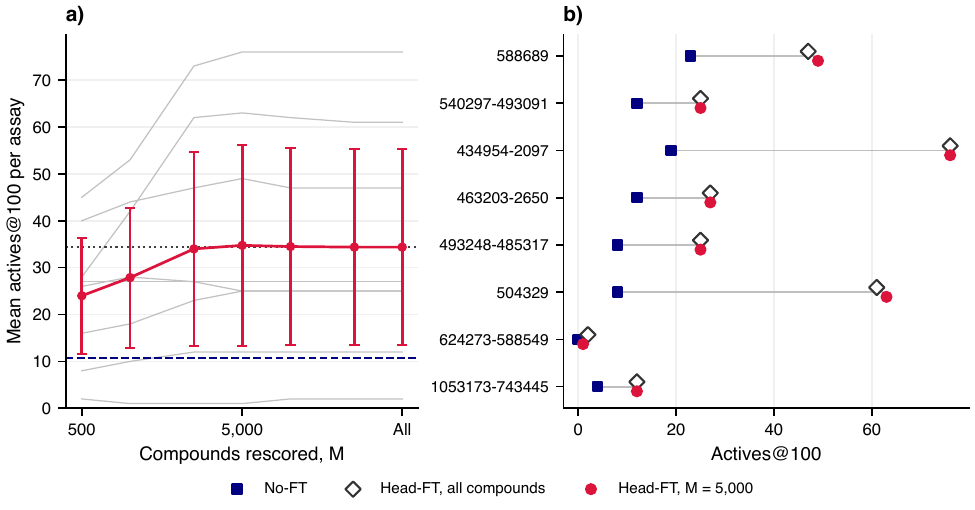}
\caption{Rescoring budget and active-compound recovery ($N=300$, $k=100$). (a) Gray lines show per-target values; red points and error bars show arithmetic means and 95\% confidence intervals. Horizontal dashed and dotted lines show mean No-FT and full head-FT recovery, respectively. All denotes the complete evaluation sets, plotted at their mean size across targets. (b) Per-target recovery for No-FT, rescoring 5,000 compounds, and full rescoring.}
\label{fig:cascade}
\end{figure}

To examine whether hit recovery can be improved without rescoring the complete evaluation set, we restricted scoring after fine-tuning to the top compounds in the initial ranking.
We reranked the top $M$ No-FT compounds using head-FT and selected the top $k$ as final candidates. We varied the rescoring budget $M$ and compared the number of actives among the final candidates with No-FT selection without rescoring and with full rescoring.
Figure~\ref{fig:cascade} shows the results for fine-tuning with 300 labels and selecting 100 final candidates ($N=300$, $k=100$).

Rescoring the top 5,000 compounds increased the mean number of actives among the 100 final candidates from 10.75 to 34.75 per target, close to the 34.38 obtained with full rescoring. Details of the paired comparisons across targets are provided in Section S2 of the Supporting Information.

The 5,000 compounds represent approximately 10\% of each evaluation set. Thus, narrowing the candidate set using the initial ranking and then selecting compounds with the fine-tuned head reduced the number of compounds requiring rescoring by approximately 90\% while yielding hit recovery close to that of full rescoring.
With $k=100$ final candidates, rescoring the top 5,000 compounds also yielded recovery close to that of full rescoring at $N=40$ and 100 (Figure S1 of the Supporting Information).

\subsection{Effects of Fine-Tuning on Compounds Structurally Dissimilar to the Training Set}

We examined whether the benefits of fine-tuning extended to compounds with scaffolds absent from the training set or low chemical similarity to the training compounds.
Stratifying the evaluation set at $N=300$ by scaffold overlap with the training set and chemical similarity to the training actives showed that head-FT achieved higher aggregate AP than No-FT in both unseen-scaffold and low-similarity strata (Table~\ref{tab:chemical_strata}).
Compound and active counts for each stratum are provided in Section S4 of the Supporting Information.

\begin{table}[!htbp]
\centering
\caption{AP stratified by scaffold overlap and chemical similarity ($N=300$).}
\label{tab:chemical_strata}
\small
\begin{tabular}{lrrl}
\toprule
Evaluation stratum & No-FT & Head-FT & AP ratio [95\% CI] \\
\midrule
All compounds & 0.052 & 0.134 & 2.137 [1.438, 3.174] \\
Unseen scaffolds & 0.043 & 0.101 & 2.031 [1.316, 3.133] \\
Unseen generic scaffolds & 0.043 & 0.092 & 1.898 [1.245, 2.895] \\
Low similarity & 0.028 & 0.044 & 1.406 [1.012, 1.953] \\
\bottomrule
\end{tabular}

\end{table}

To examine whether these gains also translated into unseen-scaffold recovery among the final candidates, we compared the top 100 compounds from the complete evaluation sets (Table~\ref{tab:scaffold_recovery}).
Head-FT recovered more actives with unseen scaffolds than No-FT, with a target-paired mean difference of 14.50 and a 95\% confidence interval of $[0.38,\,28.62]$.
The benefits of fine-tuning therefore extended to ranking and hit recovery for compounds with scaffolds absent from the training set.

\begin{table}[!htbp]
\centering
\caption{Active-compound and unseen-scaffold recovery in the top 100 compounds ($N=300$).}
\label{tab:scaffold_recovery}
\small
\begin{tabular}{lrrr}
\toprule
Method & Actives & Unseen-scaffold actives & Unseen scaffolds \\
\midrule
No-FT & 10.75 & 7.63 & 7.50 \\
Head-FT & 34.38 & 22.13 & 15.75 \\
Nearest active & 34.75 & 13.75 & 11.38 \\
\bottomrule
\end{tabular}

\end{table}

\section{Conclusions}

We showed that updating the affinity heads with a small set of binary activity labels improves early enrichment in Boltz-2-based hit discovery.
Across eight MF-PCBA targets, fine-tuning with 300 labels increased AP and EF@1\% to 2.14 and 1.77 times their No-FT values, respectively.
The higher aggregate AP than DrugCLIP projection-head fine-tuning and LightGBM trained on the same labels demonstrated the value of incorporating assay measurements directly into Boltz-2 scoring.

The benefits extended to compounds with scaffolds absent from the training set and low similarity to the training actives, and recovery of active compounds with unseen scaffolds also increased.
Restricting rescoring to approximately the top 10\% of the initial ranking retained hit recovery close to that obtained with full rescoring.
Together, these results show that combining candidate selection by a pretrained model with reranking informed by target-assay measurements can improve hit recovery from candidate sets that include structures distinct from the training compounds.

The evaluation was limited to classification with binary activity labels from the top of the initial ranking and retrospective validation using existing measurements for eight targets.
Future work should establish effectiveness under different training-set selection strategies and label compositions and provide prospective experimental validation on new targets and compounds.
Regression fine-tuning of Boltz-2 using measured activity values has already been reported~\cite{Amini2026-ee}.
Future work could also incorporate transfer learning using labels from absolute binding free energy (ABFE) calculations~\cite{Metcalf2024-qz} to explore applications in quantitative affinity prediction and lead optimization.

Fine-tuning the Boltz-2 affinity heads with a small set of binary activity labels allows measurements from an initial screen to improve subsequent candidate prioritization.
This approach adapts a pretrained structure-based predictor to the target assay, providing a means of improving early enrichment in hit discovery using limited experimental information.

\section{Data and Software Availability}

The compound sets and labels for the eight targets are available in \texttt{mf-pcba\_test.zip},
distributed with the Boltzina v1.0.1 release (\url{https://github.com/ohuelab/boltzina/releases/tag/v1.0.1}).
The reference repository for the code and analysis results is \url{https://github.com/ohuelab/BoltzFT}.

\begin{suppinfo}
Supporting information is available free of charge.
\begin{itemize}
  \item {\tt supporting.pdf}: Per-target absolute performance, two-stage selection across all label budgets, analyses of internal representations and rank changes, chemical-stratum evaluation and hit recovery, and hyperparameter selection for the comparator methods.
\end{itemize}
\end{suppinfo}

\section*{Author Information}
\subsection*{Corresponding Author}
Masahito Ohue - School of Computing, Institute of Science Tokyo, Yokohama, Kanagawa 226-8501, Japan.\\
https://orcid.org/0000-0002-0120-1643; Email: ohue@comp.isct.ac.jp

\subsection*{Author Contributions}
K.F. and M.O. conceived the study. K.F. developed the methods and software, performed the analyses, and drafted the manuscript. M.O. contributed to interpretation and supervised the study. Both authors revised and approved the manuscript.

\subsection*{Funding}
This study was supported by Japan Science and Technology Agency (JPMJAX25LB and JPMJFR216J), Japan Society for the Promotion of Science (JP24KJ1091, JP26H02544, JP26K23906, JP23H04880, and JP23H04887), and Japan Agency for Medical Research and Development (JP26ama121026).

\begin{acknowledgement}
This study was carried out using the TSUBAME 4.0 supercomputer at Institute of Science Tokyo.
The authors used ChatGPT (OpenAI) for discussion, translation, and proofreading during manuscript preparation. All AI-assisted content was carefully reviewed and revised by the authors, who take full responsibility for the accuracy and content of the manuscript.
\end{acknowledgement}

\subsection*{Conflict of Interest}
The authors declare no competing financial interests.

\bibliography{main,additional}

\includepdf[pages=-]{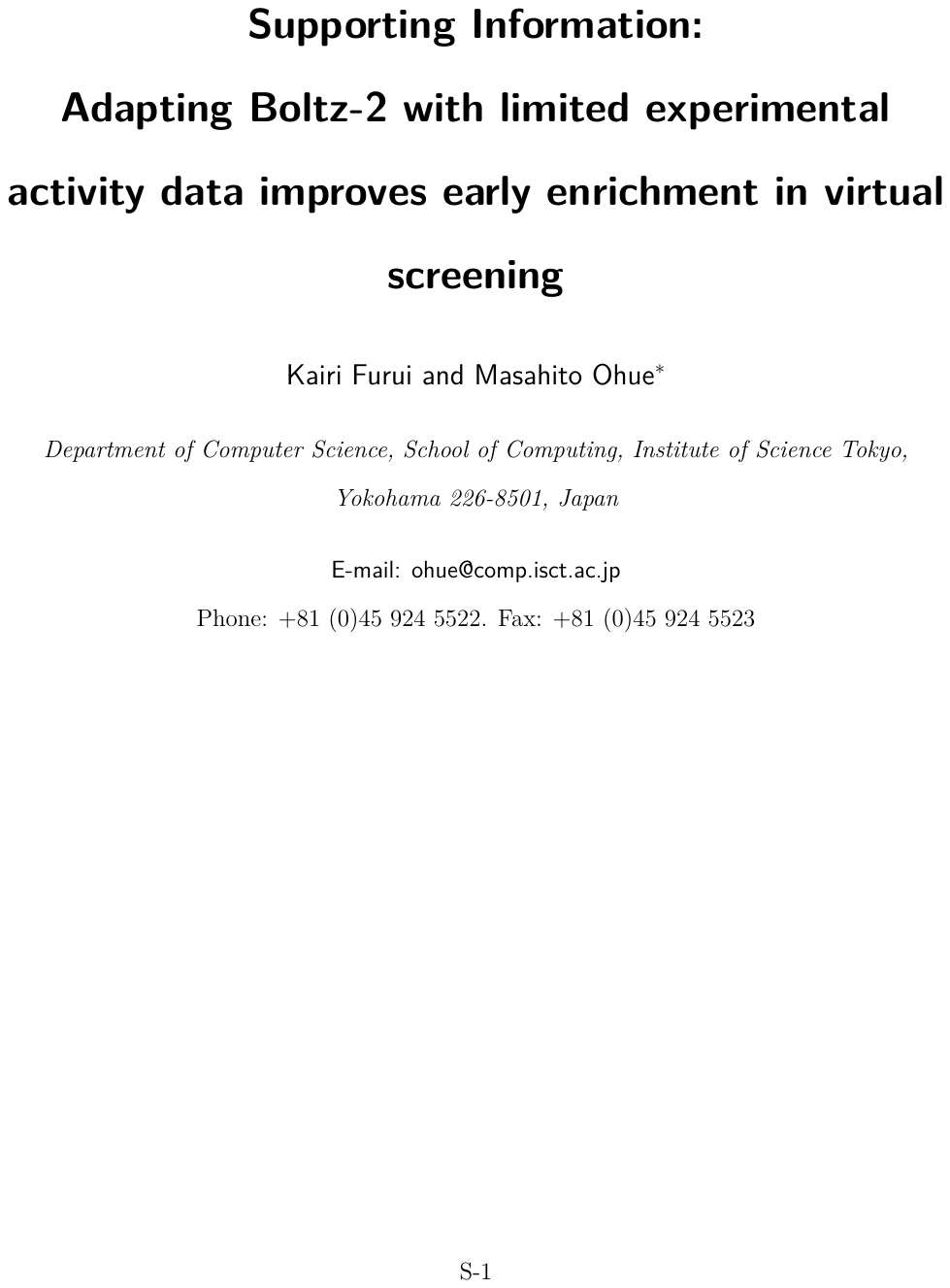}

\end{document}